\documentclass[%
 reprint,
 amsmath,amssymb,
 aps,
]{revtex4-1}
\usepackage{graphicx}
\usepackage{dcolumn}
\usepackage{bm}
\usepackage{graphicx}
\usepackage{dcolumn}
\usepackage{bm}
\usepackage{bigints}
\usepackage{color}
\usepackage{amsmath,amssymb,graphicx}
\usepackage{float}
\usepackage{braket}
\usepackage{tabularx}
\usepackage{tabulary}
\usepackage{tabu}
\usepackage{booktabs}

\newcolumntype{C}{>{\centering\arraybackslash}X}

\begin{document}
\preprint{APS/123-QED}
\title{Supersymmetric Analysis of Stochastic Micro-Bending in Optical Waveguides}

\thanks{This research is supported by grant number W911NF-19-1-0352 from the United States Army Research Office.}%

\author{Stuart Ward$^{1,2}$}
\author{Rouzbeh Allahverdi$^1$}%
\author{Arash Mafi$^{1,2}$}
\email{mafi@unm.edu}
\affiliation{$^1$Department of Physics \& Astronomy, University of New Mexico, Albuquerque, NM 87106, USA.\\
$^2$Center for High Technology Materials, University of New Mexico, Albuquerque, New Mexico 87106, USA.
}

\date{\today}
\begin{abstract}
Micro-bending attenuation in an optical waveguide can be modeled by a Fokker-Planck equation. It is shown that a supersymmetric transformation applied to the Fokker-Planck equation is equivalent to a change in the refractive index profile, resulting in a larger or smaller attenuation. For a broad class of monomial index profiles, it is always possible to obtain an index profile with a larger micro-bending attenuation using a supersymmetric transformation. However, obtaining a smaller attenuation is not always possible and is restricted to a subset of index profiles.  
\end{abstract}

\maketitle

\section{Introduction}
Supersymmetric Quantum Mechanics (SUSY-QM) was first introduced as a toy model to study SUSY breaking~\cite{Witten1}, and has since evolved into a powerful theoretical framework with many applications in theoretical physics~\cite{Witten2,Cooper1,Cooper2,Cooper3,Chumakov,Junker}. Such application include the classification of potentials with identical spectra~\cite{Jafarizadeh}, analytical derivation of spectra in certain classes of quantum Hamiltonians~\cite{Dutt}, extensions to the WKB approximation~\cite{Dutt2}, analysis of the Fokker-Planck equation (FPE) in statistical mechanics~\cite{Junker}, and random matrix theory~\cite{Zirnbauer}, disorder, and chaos~\cite{Efetov}. Using the close analogy between the Schr\"{o}dinger equation for a particle in a potential and light propagation in a dielectric medium, SUSY-QM has also been employed extensively in various optical waveguide designs and optical scattering problems over the past few years~\cite{Ganainy2012,Miri2013,Miri2013-2,Heinrich2014,Heinrich2014-3,Miri2014,Ganainy2015,Walasik2018,Midya2018,Midya2019,Walasik2019,Zhong2019,Hokmabadi2019}.

Another application of SUSY-QM is in the analysis of classical stochastic dynamics with one Cartesian degree of freedom~\cite{PaSo79,PaSo82,Junker}. 
It uses the fact that the FPE can be mapped into an imaginary-time Schr\"{o}dinger equation
\begin{align}
\label{eq:ITSE}
-\dfrac{\partial}{\partial t}\Psi_\pm(t,x)=H_\pm\Psi_\pm(t,x),
\end{align}
where $H_\pm=-\partial^2_x+W^2(x)\pm W^\prime(x)$ are the partner Hamiltonians in the Witten's model, and $W(x)$ is the superpotential. The eigenvalues of $H_\pm$ are the decay rates associated with the eigenfunctions of the imaginary-time Schr\"{o}dinger equation. In the presence of an unbroken SUSY, $H_+$ and $H_-$ have identical positive eigenvalues (positive decay rates), while $H_-$ has an additional vanishing eigenvalue, corresponding to its ground-state. In this paper, we explore how this formalism can be used to solve practical problems in classical stochastic optics. In particular, we consider the problem of micro-bending in optical waveguides, which is an important issue affecting the optical communications networks~\cite{Gardner,Jin}. Undesirable fiber-fiber and fiber-cable interactions, caused by manufacturing of fiber cables or thermal-induced variations, result in random microscopic bends leading to signal loss. The estimation of microbending loss has been performed in various publications. In particular, Jin and Payne~\cite{Jin} have presented a detailed numerical investigation, based on an analytical model, and have validated the model by comparing with experimental measurements. Therefore, there is a good coverage on the estimation of microbending loss in the existing literature. Our intention here is not to present a new way of estimating microbending attenuation; rather, this paper is framed around using SUSY-QM to design an optical waveguide with a reduced micro-bending loss. Moreover, the overall formalism and ideas can be readily applied to a wide range of problems involving stochastic dynamics.
\section{Optical waveguide with micro-bending}
An optical waveguide is a transversely inhomogeneous structure, usually containing a region of increased refractive index, compared with the surrounding medium. Here, to illustrate the main ideas, we consider the micro-bending of a planar waveguide structure. Generalization to an optical fiber is possible but is beyond the scope of the present work. We note that intrinsic fluctuations in the refractive index, including those stemming from random glass density fluctuations or the presence of impurities, lead to unwanted back-scattering and mode mixing. For the case of micro-bending, the refractive index fluctuation is not intrinsic but can be modeled as a modification of the refractive index by the addition of a linear term to the first order in the radial coordinate over the radius of the curvature. Such a tilt in the refractive index profile results in leaking radiation. 

For the problem of interest, we assume a planar waveguide structure defined by an inhomogeneous refractive index $n(y)$, which is independent of the $x$ and $z$ coordinates. The light rays are confined in the $y$ direction by the inhomogeneous refractive index, while propagating freely in the $z$ direction. The trajectories of light rays are determined by Fermat's principle~\cite{SalehTeich}, $\delta\int^b_an(y)ds=0$, where $ds$ is the differential length on a trajectory between points $a$ and $b$ in the $y-z$ plane. Using the calculus of variation results in the following differential equation for the ray trajectory $y(z)$:
\begin{align}
\dfrac{d}{ds}\left(n\dfrac{dy}{ds}\right)
=\dfrac{\partial n}{\partial y}~~~~~~,\qquad
ds=\sqrt{dy^2+dz^2}.
\end{align}

To model the micro-bending, we assume that the optical waveguide is bent in the $y$-$z$ plane with a local curvature of $\kappa(z)$. A bent waveguide can be conformally mapped to a straight waveguide, where the effect of the local bending can be modeled as a variation in the refractive index~\cite{Heiblum,MafiAcoustic} given by $n(y,z)\approx n_{s}(y)\left[1-\,\kappa(z)\,y\right]$. $n_{s}$ is the effective refractive index for the straight waveguide (without micro-bending). In the paraxial approximation, the ray trajectory is almost parallel to the $z$ axis, so we can assume $ds\approx dz$. Moreover, the maximum of the refractive index is commonly at or near the waveguide center at $y=0$, and the refractive index gradually decreases away from the center. Taking these points into consideration and in the absence of large transverse variations, the ray propagation equation can be simplified into the dynamical equation of a point particle subject to a net force of $F(y)-\kappa(z)$, where the conservative force $F(y)$ is exclusively due to the inhomogeneous refractive index $n_s(y)$~\cite{Rousseau,Arnaud}:
\begin{align}
\label{eq:rayNewton}
\dfrac{d^2y}{dz^2}\approx F(y)-\kappa(z).
\end{align}
Note that the longitudinal coordinate $z$ plays the role of time for the point particle. The conservative force $F(y)$ is related to the effective potential $U(y)$ by
\begin{align}
F(y)=-dU/dy,\qquad U(y)=1-n_s(y)/n_0,
\end{align}
where $n_0=n_s(y=0)$ and $U(0)=0$. Because the refractive index decreases away from the center of the waveguide, the potential increases at large $|y|$ and $U(y)$ is a confining potential. 

We assume that the waveguide curvature $\kappa(z)$ is a zero mean ($\braket{\kappa(z)}=0$) white stochastic process with the autocorrelation given by
\begin{align}
\label{eq:corr}
\braket{\kappa(z)\kappa(z')} = \kappa_{0}^2\delta(z-z').
\end{align}
The ray propagation equation~(\ref{eq:rayNewton}) can be expressed as a stochastic Langevin equation in the following form:
\begin{align}
\label{eq:langevin}
\dfrac{dy}{dz} = \theta(z),\qquad
\dfrac{d\theta}{dz} = F(y) - \kappa(z),
\end{align}
where $\theta(z)$ is the local angle of the ray relative to the $z$ axis. Instead of solving Eq.~(\ref{eq:langevin}) for each random realization and then averaging the results, it is more convenient to use the corresponding FPE, from which we can determine the probability $P$ of an optical ray to be at lateral position $y$ and angle $\theta$ at point $z$ along the fiber~\cite{Rousseau,Arnaud}:
\begin{align}
\label{eq:FokkerPlanck1}
\left[\dfrac{\partial}{\partial z}
+\theta\dfrac{\partial}{\partial y}
+F\dfrac{\partial}{\partial\theta}
-\dfrac{\kappa_0^2}{2}\dfrac{\partial^2}{\partial\theta^2}\right]
P(y,\theta,z)=0.
\end{align}
The FPE (\ref{eq:FokkerPlanck1}) describes many stochastic problems in the phase-space; therefore, the following results can be applied to a wide range of physical systems beyond the micro-bending problem that is analyzed here.

This can be further simplified by noting that the Hamiltonian of the conservative system (in the absence of stochastic noise) is given by
\begin{align}
E=\dfrac{1}{2}\theta^2+U(y),
\label{eq:Hamiltonian}
\end{align}
where $\theta$ plays the role of the velocity of the point particle with the mass $m=1$. 
In a typical graded-index optical waveguide, where the refractive index peaks in the center of the waveguide and gradually decreases away from the center, light rays follow sinusoidal-like paths along the waveguide. In the center where $U$ is zero, the angle $\theta$ and the kinetic term assume their maximum values. As a ray reaches its maximum transverse separation from the center and bends back towards the center, $\theta$ and the kinetic term become zero and $U$ reaches its maximum value equal to $E$. In other words, $E$ determines the maximum ray angle at the center and also the maximum distance a ray can reach away from the center. Of course, if the energy $E$ of a light ray is too large, it can breach the potential barrier into free space.  
In the modal language of wave optics, the energy $E$ can be related to the mode number, where a larger $E$ corresponds to a higher order mode and the energy at which a light ray breaches the potential barrier corresponds to the highest order mode supported by the waveguide.
In Eq.~(\ref{eq:FokkerPlanck1}), the microbending attenuation appears as the exponential decay of the probability $P(y,\theta,z)$ with respect to the propagation distance $z$ as will be discussed in detail later in the paper.

The phase-space action, which is a conserved quantity, is the area enclosed in the phase-space by the ray trajectory and is given as
\begin{align}
\label{eq:defIE}
I(E)=\iint d\theta dy=\oint_T\sqrt{2[E-U(y)]}dy=T\overline{\theta^2},
\end{align}
where $T$ is the ray period and overbar denotes the average over one ray period. The derivative of $I$ with respect to $E$ also takes the simple form of
\begin{align}
\label{eq:IprimeE}
I^\prime(E)=\oint_T\dfrac{1}{\sqrt{2(E-U)}}dy=T.
\end{align}
Using this information, the FPE in~(\ref{eq:FokkerPlanck1}) can be expressed in a much simpler form, where $y$ and $\theta$ are swapped with a single variable $E$ to obtain

\begin{align}
\label{eq:FokkerPlanck2}
\dfrac{2}{\kappa^2_0}\dfrac{\partial P(E,z)}{\partial z}=
-\dfrac{\partial P(E,z)}{\partial E}
+\dfrac{\partial^2[\mathbb{Z}(E)P(E,z)]}{\partial E^2},
\end{align}
where $P(E,z)$ is the probability of the ray to have energy $E$ at 
point $z$ along the waveguide, and $\mathbb{Z}(E)=I(E)/I^\prime(E)$
is twice the average kinetic energy over one ray period. 

We now make a change of variable (both in $E$ and $P$) and transform the FPE in~(\ref{eq:FokkerPlanck2}) to an imaginary-time Schr\"{o}dinger equation, similar to Eq.~(\ref{eq:ITSE}). We write $E=E(\mathcal{E})$ and $P(E,z)=B(E)Q(E,z)$ and choose the function $E(\mathcal{E})$ and the form of $B(E)$ to satisfy
\begin{align}
\mathbb{Z}{\mathcal{E}}^{\prime2} = 1,\ \mathbb{Z}=\dot{E}^2,\qquad 
2\mathbb{Z}B^\prime+\big(\dfrac{3}{2}\mathbb{Z}^\prime-1\big)B=0,
\end{align}
where ${\prime}$ and $\cdot$ denote differentiation with respect to $E$ and $\mathcal{E}$, respectively. The solutions to these equations are formally given by
\begin{align}
\label{eq:EeZ}
&\mathcal{E}=\int \dfrac{dE}{\sqrt{\mathbb{Z}}},\qquad
E=\int\sqrt{\mathbb{Z}}\, d\mathcal{E},\\ 
\nonumber
&B=\mathbb{Z}^{-3/4}\exp\Big(\int\dfrac{dE}{2\mathbb{Z}}\Big).
\end{align}
Using these changes of variables, we arrive at the imaginary-time Schr\"{o}dinger equation
\begin{align}
\label{eq:SEQ}
-\dfrac{\partial Q}{\partial Z}=
-\dfrac{\partial^2 Q}{\partial \mathcal{E}^2}
+V(\mathcal{E})Q,
\end{align}
where $Z=\kappa_0^2z/2$ and 
\begin{align}
\label{eq:VFE}
V(\mathcal{E})=\mathbb{F}^{-1} \dfrac{\partial^2}{\partial\mathcal{E}^2} \mathbb{F},
\qquad 
\mathbb{F}^4
=\big[I^2(E)\big]^\prime
=2\big[\dot{I}(\mathcal{E})\big]^2.
\end{align}

We now have all the required ingredients to make a SUSY transformation on Eq.~(\ref{eq:SEQ}). Once we find the SUSY partner potential to $V(\mathcal{E})$, we can work our way backward to find the corresponding $I(E)$ from Eq.~(\ref{eq:VFE}) and consequently $Z(E)$, from which we can find $U(y)$. This will be achieved by using the Abel transform-pairs, which can be expressed as  
\begin{align}
\label{eq:IprimeE2}
I^\prime(E)&=\int^E_0\left(\dfrac{2\sqrt{2}}{\partial_yU}\right)\dfrac{dU}{\sqrt{E-U}},\\
\label{eq:invIprimeE2}
\dfrac{2\sqrt{2}}{\partial_yU}&=\frac{1}{\pi} \frac{\partial}{\partial U} \int_{0}^{U} I^\prime(E) \frac{d E}{\sqrt{U-E}}.
\end{align}
Note that Eq.~(\ref{eq:IprimeE2}) is essentially the same as Eq.~(\ref{eq:IprimeE}), albeit with a change of variable. Equation~(\ref{eq:invIprimeE2}) gives us the required information about the new $U(y)$. In other words, the Abel transform-pairs allow us to relate the phase-space action to the refractive index profile of the optical waveguide and vice versa.
\vskip 2mm
\noindent
{\em Example of a monomial potenial:} As a concrete example we explore a monomial potential of the form $U(y)=\Delta |y/y_c|^\alpha$ in some detail. This model arises in graded-index optical waveguides~\cite{Gloge,MafiGraded}, where $y_c$ is the waveguide half-width, $\alpha$ controls the shape of the index profile, and $\Delta$ determines the refractive index contrast of the center relative to the core edges at $y=\pm y_c$. Note that it is common to take $\alpha\approx 2$ in practical optical waveguides~\cite{Gloge,MafiGraded}, because it reduces the modal disperion in optical waveguides by equalizing the group velocities of the modes. In the ray picture, for $\alpha\approx 2$, all rays of different energy $E$ (and maximum value of $\theta$) reach the end of the waveguide at the same time. 

In the following discussion, because the potential is assumed to be symmetric under $y\to -y$, we only consider the $y\ge 0$ region without loss of generality. For this problem, we can evaluate $I(E)$ using 
\begin{align}
&I(E)=\int^E_0\left(\dfrac{4\sqrt{2}}{\partial_yU}\right)\sqrt{E-U}dU\\
\nonumber
&=\dfrac{\sqrt{8\pi}\, \Gamma(1+1/\alpha)}{\Gamma(3/2+1/\alpha)}
y_c\,\Delta^{-1/\alpha}E^{1/\zeta},\ 
{\rm where} \ \zeta=\dfrac{2\alpha}{\alpha+2}.
\end{align}
In this special case, the ratio $\mathbb{Z}(E)$ takes the simple form of
$\mathbb{Z}(E)=\zeta E$, and we obtain $4E=\zeta\mathcal{E}^2$ and $B^{2\alpha}\propto E^{(1-\alpha)}$. Using these results and Eq.~(\ref{eq:VFE}), we obtain:
\begin{align}
\label{eq:Vsolution}
V(\mathcal{E})
=\dfrac{1-\alpha}{\alpha^2\mathcal{E}^2}
=\dfrac{\nu^2-1/4}{\mathcal{E}^2},\ 
{\rm where} \ \nu=\dfrac{2-\alpha}{2\alpha}.
\end{align}
Before we present the solution to Eq.~(\ref{eq:SEQ}), we note that lossy rays are those which can breach the potential barrier into the free space, i.e. with an energy larger than a given $\mathbb{E}$. This results in a boundary condition of $P(\mathbb{E},z)=0$, or equivalently $Q(\boldsymbol{\epsilon},Z)=0$, where $4\mathbb{E}=\zeta\boldsymbol{\epsilon}^2$. Attempting an eigensolution of the form $Q(\mathcal{E})=Q_m(\mathcal{E})e^{-\lambda_m Z}$, we obtain
\begin{align}
\label{eq:Qm}
Q_m(\mathcal{E})=\sqrt{\mathcal{E}/\boldsymbol{\epsilon}}\,J_\nu(\sqrt{\lambda_m}\mathcal{E}),
\quad
\lambda_m=u^2_{\nu m}/\boldsymbol{\epsilon}^2,
\end{align}
where $J_\nu$ is the Bessel function of order $\nu$ and $u_{\nu m}$ is the $m$th zero of $J_\nu$, where $m=0, 1, \cdots$. Recall that $Z=\kappa_0^2z/2,$ where $\kappa_0$ is defined through the autocorrelation of the curvature which is defined in Eq.~(\ref{eq:corr}). The relationship between attenuation and curvature can now be seen from the term $e^{-\lambda_m Z}$ of the above eigensolution. Furthermore, recall that the goal of performing the SUSY transformation is to find a new potential which possesses an additional eigensolution with a decay constant that is smaller than the smallest decay constant ($\lambda_0$) of the above eigensolution.

To perform a SUSY transformation, we need to calculate the superpotential
$W=-\dot{Q}_0/Q_0$, where $Q_0$ is the eigensolution with the smallest decay coefficient, i.e. $\lambda_0$. We obtain
\begin{align}
\label{Eq:W-1}
W = -\dfrac{\nu + 1/2}{\mathcal{E}}+\sqrt{\lambda_{0}}\dfrac{J_{\nu+1}(\sqrt{\lambda_{0}}\mathcal{E})}{J_{\nu}(\sqrt{\lambda_{0}}\mathcal{E})}.
\end{align}
The partner potentials are derived from $V_{\pm} = W^2 \pm \dot{W}$. We obtain
\begin{align}
\label{Vm1}
&V_{-} = -\lambda_{0}+\dfrac{\nu^2-1/4}{\mathcal{E}^2},\\
\label{Vp1}
&V_{+} = +\lambda_0+\dfrac{\mathcal{G}}{\mathcal{E}^2},\\
\nonumber
&\mathcal{G}=\dfrac{3}{4}+\nu(\nu+2)-\dfrac{(4\nu+2)\sqrt{\lambda_0}\mathcal{E}J_{\nu+1}}{J_{\nu}}
+\dfrac{2\lambda_0\mathcal{E}^2J^2_{\nu+1}}{J^2_{\nu}},
\end{align}
where we have suppressed the argument of the Bessel functions, all being $\sqrt{\lambda_{0}}\mathcal{E}$. Note that $V_-$ is the same as the potential in Eq.~(\ref{eq:Vsolution}) except it is shifted by $\lambda_0$ as expected~\cite{Suk85b}. 
We also have $\lambda_0=u^2_{\nu0}/\boldsymbol{\epsilon}^2$; therefore, we are only interested in $V_{\pm}$ in the range of $0\le \mathcal{E}< \boldsymbol{\epsilon}$, where $V_+$ blows up at the upper boundary. 

\begin{figure}[h]
\begin{center}
\includegraphics[width=0.35\textwidth]{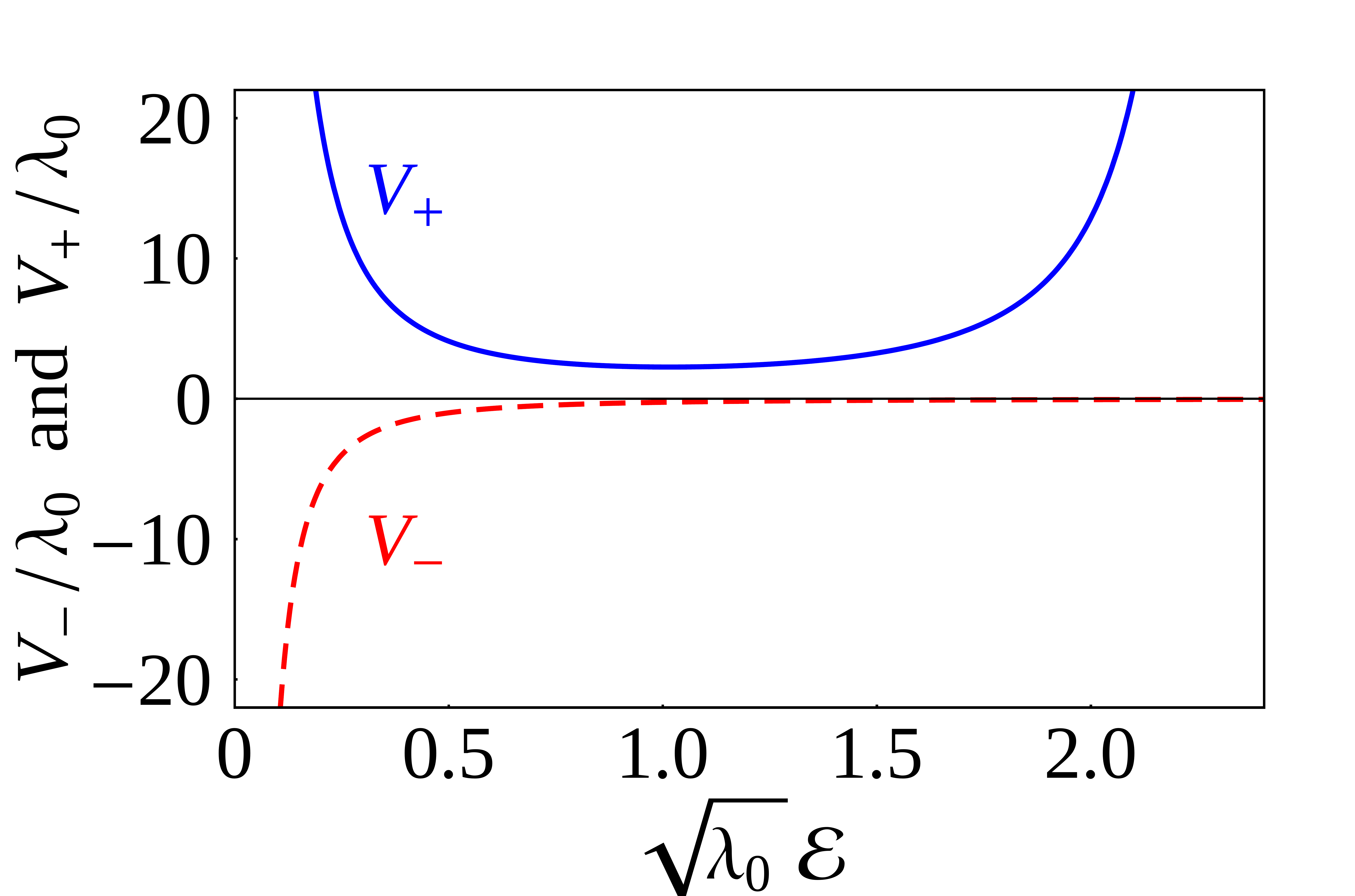}
\caption{$V_{\pm}$ from Eqs.~(\ref{Vm1}) and~(\ref{Vp1}) are plotted as a function of $\sqrt{\lambda_{0}}\mathcal{E}$ for $\nu=0$. The potentials are shifted up by $\lambda_0$ and are expressed in units of $\lambda_0$.}
\label{fig:Vplot}
\end{center}
\end{figure}

To establish a concrete numerical example, we consider the case of a quadratic graded index with $\alpha=2$ and $\nu=0$, where $u_{00}\approx 2.405$. We assume that those rays that can make it to $y\ge y_c$ are lost to free space, resulting in $\mathbb{E}=\Delta$ or equivalently $\boldsymbol{\epsilon}=2\sqrt{\Delta}$. We also assume $\Delta=0.01$, which is a reasonable value in graded index optics, resulting in $\boldsymbol{\epsilon}=0.2$ and $\lambda_0\approx 144.58$. In this case, we plot $V_{\pm}$ from Eqs.~(\ref{Vm1}) and~(\ref{Vp1}), except we redefine these functions by adding $\lambda_0$ to get the original $V_{-}$ back. These shifted potentials are plotted in FIG.~\ref{fig:Vplot}.    
In physical units, the lowest attenuation coefficient corresponds to $\lambda_0\kappa_0^2/2$. Considering a waveguide with $y_c\approx 1$\textmu m and $\kappa^2_0\approx 4.8\times 10^{-6}$\textmu m$^{-1}$, which corresponds to a fluctuation in radius of curvature of 457\textmu m over a longitudinal distance comparable to $y_c$, the lowest attenuation is obtained to be around 3dB/mm. Less fluctuation corresponds to larger curvature radii fluctuating at a slower rate with the propagation distance.

\begin{figure}[h]
\begin{center}
\includegraphics[width=0.48\textwidth]{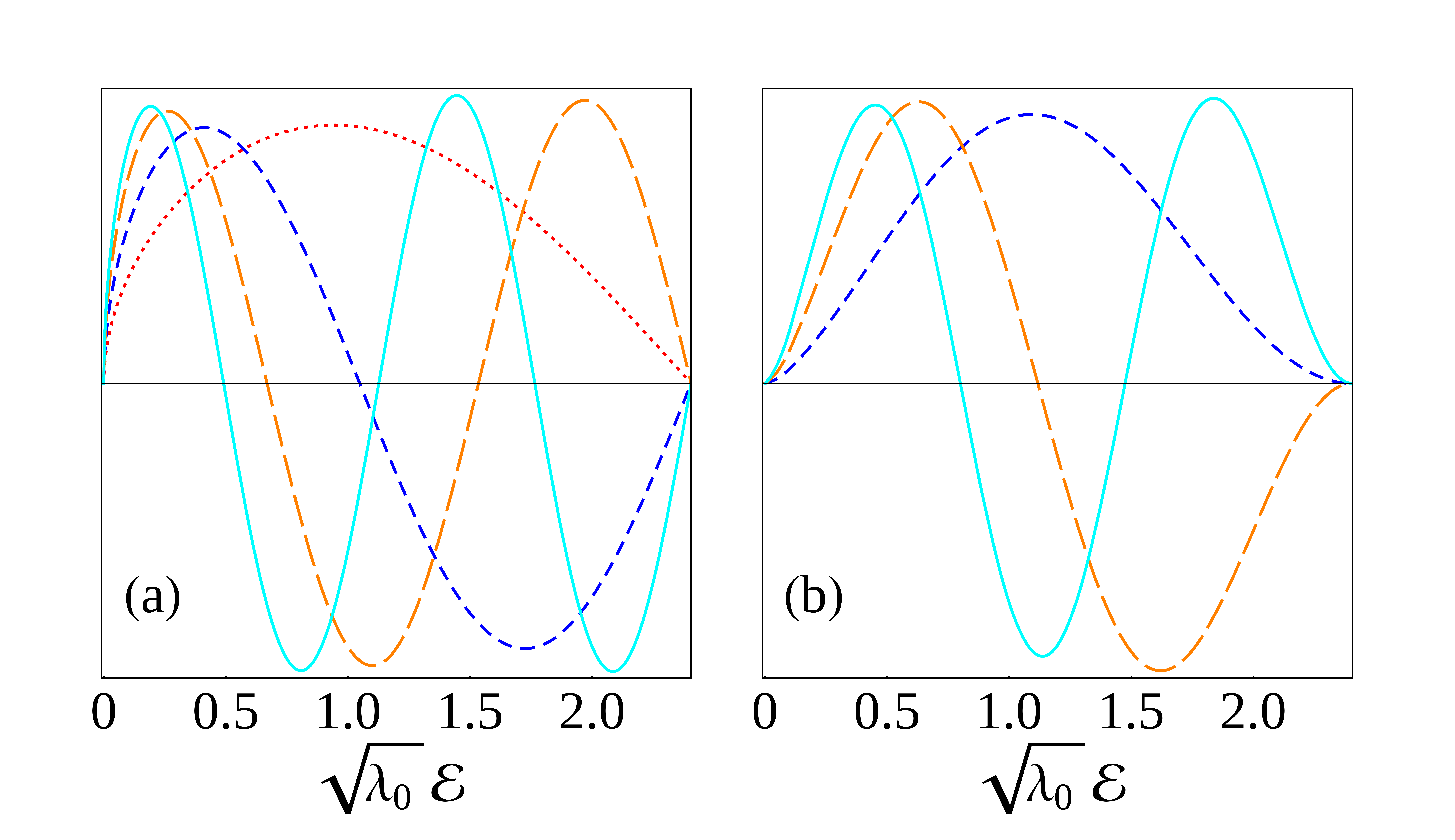}
\caption{(a) The lowest 4 eigenfunctions $Q^-_m$ ($m=0,1,2,3$) corresponding to $V_-$, and (b) the lowest 3 eigenfunctions $Q^+_m$ ($m=1,2,3$) corresponding to $V_+$ are plotted.}
\label{fig:wavefunctions}
\end{center}
\end{figure}

To verify that $V_+$ is a true superpartner potential to $V_-$, we solve the eigenvalue problem with $V_+$ in Eq.~(\ref{eq:SEQ}) assuming the $Z$-dependence of the form $e^{-\lambda_m Z}$ and the Dirichlet boundary condition: $Q^+(0)=Q^+(\boldsymbol{\epsilon})=0$ (superscript $+$ signifies its correspondence to $V_+$). The eigenvalues must match those of the excited states obtained from $V_-$. We plot, in FIG.~\ref{fig:wavefunctions}(a), the lowest 4 eigenfunctions $Q^-_m$ ($m=0,1,2,3$) corresponding to $V_-$ from Eq.~(\ref{eq:Qm}), where $m$ corresponds to how many times the eigenfunction crosses the $Q=0$ horizontal axis. The corresponding eigenvalues are $\lambda_m=(u_{0m}/u_{00})^2\lambda_0$ for $m=0,1,2,3$, respectively. Similarly, in FIG.~\ref{fig:wavefunctions}(b), we plot the lowest 3 eigenfunctions $Q^+_m$ ($m=1,2,3$) corresponding to $V_+$ that are evaluated numerically, where $m-1$ corresponds to the number of crossings with the $Q=0$ horizontal axis. 
It can be confirmed, numerically, that the eigenvalues corresponding to $Q^+_1$, $Q^+_2$, and $Q^+_3$ match those of $Q^-_1$,
$Q^-_2$, and $Q^-_3$, respectively, as expected from an unbroken SUSY.

We now outline the procedure to evaluate $U(y)$ corresponding to $V_+$. We first insert $V_+$ from Eq.~(\ref{Vp1}) into Eq.~(\ref{eq:VFE}) and obtain a numerical solution for $\mathbb{F}$. The boundary conditions for $\mathbb{F}$ are set at $\mathcal{E}\to 0$: we make a Taylor expansion of $V_+$ near this point and analytically solve Eq.~(\ref{eq:VFE}) near $\mathcal{E}=0$ to find that $\mathbb{F}(\mathcal{E})\propto \mathcal{E}^{3/2}$ and $\dot{\mathbb{F}}(\mathcal{E})\propto (3/2)\mathcal{E}^{1/2}$. These expressions guide us to set the boundary conditions for $\mathbb{F}$ and $\dot{\mathbb{F}}$ for $\mathcal{E}\to 0$. Note that Eq.~(\ref{eq:VFE}) can determine $\mathbb{F}$ only up to an overall factor, and this is rooted in the fact that $I(E)$ in Eq.~(\ref{eq:defIE}) can be multiplied by a constant factor without changing the FPE. The next step is to use $\sqrt{2}\dot{I}=\mathbb{F}^2$, assuming that $I(0)=0$, to numerically evaluate $I(\mathcal{E})$. It can be shown that $\mathbb{Z}(\mathcal{E})=2I^2/\mathbb{F}^4$, so we use this expression to evaluate $\mathbb{Z}(\mathcal{E})$. Next, $\mathbb{Z}(\mathcal{E})$ along with the transformations in Eq.~(\ref{eq:EeZ}) are used to relate $E$ to $\mathcal{E}$. Using this relationship and $I(\mathcal{E})$, we can find $I(E)$ and consequently $I^\prime(E)$, to be used in Eq.~(\ref{eq:invIprimeE2}) to find $U(y)$. Considering that $U(y=0)=0$, we can rewrite Eq.~(\ref{eq:invIprimeE2}) in the simpler form of
\begin{align}
\label{eq:invIprimeE22}
y=\dfrac{1}{2\sqrt{2}\pi}\int_{0}^{U} I^\prime(E) \dfrac{d E}{\sqrt{U-E}}.
\end{align}

\begin{figure}[htp]
\begin{center}
\includegraphics[width=0.48\textwidth]{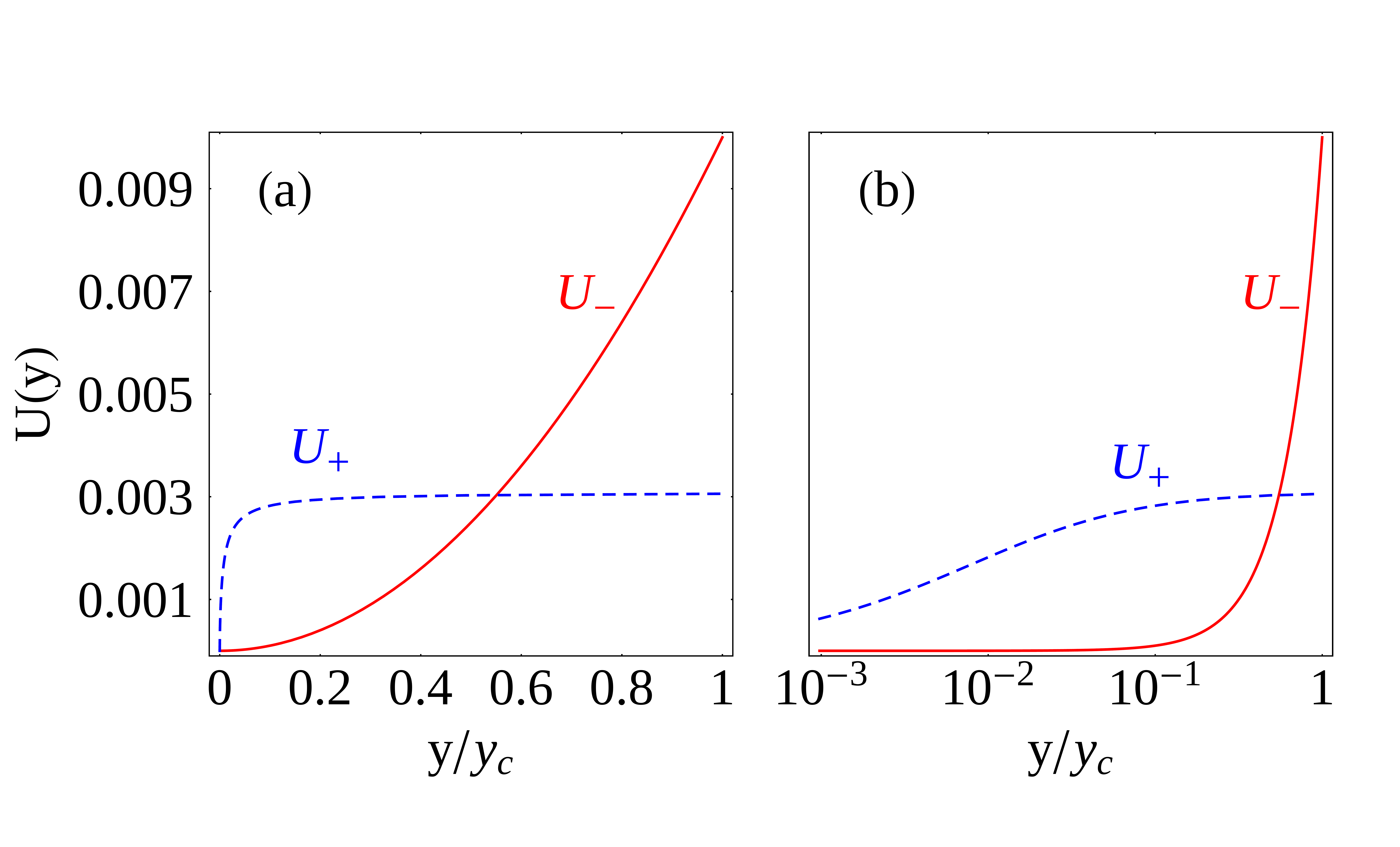}
\caption{(a) $U_-=\Delta(y/y_c)^2$ (corresponding to $V_-$) is plotted in comparison to $U_+$ (corresponding to $V_+$), where $U_+$ is obtained according to the outlined numerical procedure. Plots in (b) are the same as (a) but the vertical axis is logarithmic. Note that $V_{-,-}$ and $U_{-,-}$ do not exist for the case of $\alpha=2$ treated here.}
\label{fig:potentials}
\end{center}
\end{figure}
Because the redundant multiplicative factor mentioned earlier propagates to $I^\prime(E)$ in Eq.~(\ref{eq:invIprimeE22}), the right-hand side of Eq.~(\ref{eq:invIprimeE22}) can only be determined up to an overall factor. This ambiguity is there because nowhere in our formalism did we use the actual value of $y_c$, so we can normalize the result obtained in Eq.~(\ref{eq:invIprimeE22}) to $y_c$. The final result is shown in Fig.~\ref{fig:potentials}, where $U_-$ corresponds to $V_-$ and is given by $\Delta(y/y_c)^2${\bf ,} and $U_+$ corresponds to $V_+$ and is obtained according to the numerical procedure outlined above. We remind that because these potentials are symmetric under $y\to -y$, only the $y\ge 0$ region is shown in Fig.~\ref{fig:potentials}. Note that $U_-$ is quadratic and concave; however, $U_+$ is convex and its derivative at $y=0$ blows up. In practical situations, a slightly rounded potential at $y=0$ does not alter the behavior of the system in a notable way.

Let us recap the behavior of the two systems as $\mathcal{E}\to 0$. For $V_-$, $\mathbb{F}\propto\mathcal{E}^{1/2}$, so $\dot{I}\propto\mathcal{E}$ and $I\propto\mathcal{E}^2$. Therefore, $\mathbb{Z}\propto \mathcal{E}^2$, which results in $E\propto \mathcal{E}^2$. As such, $I\propto E$ and $I^\prime={\,const.}$, which results in $\partial_yU=0$ for $y\to 0$ according to Eq.~(\ref{eq:invIprimeE2}). On the other hand, for $V_+$, $\mathbb{F}\propto\mathcal{E}^{3/2}$, so $\dot{I}\propto\mathcal{E}^3$ and $I\propto\mathcal{E}^4$. Therefore, $\mathbb{Z}\propto \mathcal{E}^2$, which results in $E\propto \mathcal{E}^2$. As such, $I\propto E^2$ and $I^\prime\propto E$, which results in $|\partial_yU|\to \infty$ for $|y|\to 0$ according to Eq.~(\ref{eq:invIprimeE2}). The lowest order decay coefficient for $U_+$ is $\lambda_1=(u_{01}/u_{00})^2\lambda_0$, which is larger than that of $U_-$ ($\lambda_0$) and this agrees with the form of $U_+$ and the fact that it is shallower than $U_-$. 
\section{Lowering the loss}
We would now like to carry out the reverse operation and find out if there exists a SUSY-constructed potential, which we refer to as $U_{-,-}$, that can result in a lower minimum attenuation than $\lambda_0$. To achieve this, one must find a potential $V_{-,-}$ that is the superpartner potential to (a slightly shifted) $V_{-}$ but with a lower (zero energy) ground state. This implies that we need to find a $W$ that satisfies $W^2+\dot{W}=V_-$, where the positive sign behind $\dot{W}$ implies that $V_-$ is the higher energy superpartner potential, and 
$W^2-\dot{W}=V_{-,-}$. 

It is seen from Eq.~(\ref{eq:Vsolution}) that regardless of a finite shift in $V_-$, the behavior of $V_-$ for $\mathcal{E}\to 0$ is of the form $(\nu^2-1/4)/\mathcal{E}^2$. Previously, setting $W^2-\dot{W}=V_-$ resulted in $W\sim -(\nu+1/2)/\mathcal{E}$, consistent with Eq.~(\ref{Eq:W-1}) for $\mathcal{E}\to 0$ and $Q_0\sim \mathcal{E}^{1/2+\nu}$ with a Dirichlet boundary condition for $Q_0$ at $\mathcal{E}=0$. For $\alpha > 0$, $\nu>-1/2$,
and hence $Q_0$ is always well-behaved for $\mathcal{E}\to 0$, which is reassuring. Now, setting $W^2+\dot{W}=V_-$ results in $W\sim (1/2-\nu)/\mathcal{E}$. As such, the ground state of $V_{-,-}$, which would have to satisfy $W=-\dot{Q}_0/Q_0$ gives $Q_0\sim \mathcal{E}^{-1/2+\nu}$. This form of $Q_0$ is only non-diverging for $\mathcal{E}\to 0$, and hence normalizable, if $\nu>1/2$. This happens if $\alpha<1$, which implies $V_{-,-}$ can be found only if $\nu>1/2$ (equivalently $\alpha<1$). 

This clearly shows that for the quadratic potential $\Delta(y/y_c)^2$, with $\nu=0$ and $\alpha=2$, there exists no $V_{-,-}$. In other words, one cannot find a lower energy superpartner potential that results in a lower micro-bending attenuation. However, for $\Delta(y/y_c)^\alpha$ with $\alpha<1$ which is of the concave form, $V_{-,-}$ can be found. It can be used to construct a potential $U_{-,-}$, and hence a refractive index that has a lower micro-bending attenuation. The evidence of this already exists in our 
calculations. Previously, we used $\nu=0$ for $V_-$ in Eq.~(\ref{Vm1}), corresponding to $U_-=\Delta(y/y_c)^2$, to construct $V_+$ in Eq.~(\ref{Vp1}) corresponding to $U_+$. As such, one expects to be able to construct $V_-$ from $V_+$. 
To reverse the logic, we note that the behavior of $V_+$ for $\mathcal{E}\to 0$ is of the form $V_+\sim (3/4)/\mathcal{E}^2$ for $\mathcal{E}\to 0$. Because $\nu^2-1/4=3/4$ results in $\nu=1>1/2$, so it is no wonder that $V_-$ exists as the lower energy partner of $V_+$. 

Using these results, we can make general arguments on how the shape of $U$ changes under SUSY transformations, both going up ($V_-$ to $V_+$) and possibly going down ($V_-$ to $V_{-,-}$). Recall that $W^2-\dot{W}=(\nu^2-1/4)/\mathcal{E}^2$ results in $W\sim -(\nu+1/2)/\mathcal{E}$, while $W^2+\dot{W}=(\nu^{\prime 2}-1/4)/\mathcal{E}^2$ results in $W\sim (1/2-\nu^\prime)/\mathcal{E}$. Thus, for an ``up-ward'' SUSY transformation
$\nu^\prime = \nu + 1$ or, equivalently, $\alpha^\prime=\alpha/(1+\alpha)$. We have already seen this for $\nu=0$ and $\alpha=2$ in $V_-$, which resulted in $\nu^\prime=1$ and $\alpha^\prime=2/3$ in $V_+$. In fact, it can be shown that $U_+$ scales as $\sim (y/y_c)^{2/3}$ near $y=0$. The relation $\alpha^\prime = \alpha/(1 + \alpha)$ shows that in an ``up-ward'' SUSY transformation the power in the potential of the form $U \sim (y/y_c)^\alpha$ decreases and the potential becomes (more) concave. Another important point is that the ``up-ward'' SUSY transformation is always permitted~\cite{Suk85b}. On the other hand, for a ``down-ward'' SUSY transformation, i.e. for $V_-$ going down to $V_{-,-}$, we have $\alpha^\prime=\alpha/(1-\alpha)$. First, $\alpha<1$ is required for this to make sense, as we showed above, and the potential becomes (more) convex under this operation. For example, $\alpha=1/2$ in $V_-$ gives $\alpha^\prime=1$ for $V_{-,-}$.

We would like to emphasize two important points regarding the results obtained in this paper. First, we make a general observation that the attenuation can be lowered only in profiles with $\alpha<1$. In particular, for the interesting case of $\alpha\approx 2$, lowering the attenuation is not possible. However, our results are strictly true only for monomial index profiles and we have not explored other forms. Therefore, we are not ruling out the possibility that a SUSY transformation can be used to lower the attenuation for a non-monomial near-quadratic index profile. Second, our conclusions are based only on SUSY transformations and we are not ruling out the possibility of lowering the attenuation for the monomial index profile with $\alpha>1$ by leveraging other non-SUSY transformations. In practical optical waveguide designs (especially for optical fibers), some common values of $\alpha$ are $\alpha\gg 1$ for a step-index waveguide, $\alpha\approx 2$ for a graded-index waveguide to reduce the modal dispersion, and $\alpha\approx 1$ for a dispersion-shifted waveguide~\cite{Agrawal}. One can envision $\alpha<1$ profiles for dispersion shifting as well. However, considering that $\alpha<1$ is not a commonly used profile, our finding can be considered as a ``negative result'', showing that the attenuation of commonly used optical waveguides cannot be reduced by a SUSY transformation, subject to the limitations stated above.
\section{Conclusions}
We have employed SUSY-QM in the analysis of a classical stochastic optics problem: the random micro-bending loss of a waveguide. For a general class of monomial-shaped potentials (refractive index profile), we showed that a SUSY transformation can always be used to make the waveguide more lossy ($V_-$ to $V_+$), but only certain refractive index profiles are amenable to a (reverse) SUSY transformation to make the waveguide less lossy ($V_-$ to $V_{-,-}$). The results of this paper can potentially be used as a seed for the optimization or inverse design of optical waveguides. Our work may also serve as a framework to investigate similar phenomena in optics or other areas of physics described by a FPE that can be investigated by using the SUSY-QM formalism. 

We would like to emphasize that our work is performed using the geometrical ray picture in what effectively constitutes a highly multimode waveguide. We would like to acknowledge the interesting result obtained in Ref.~\cite{Grillot} on the loss due to sidewall roughness for single mode propagation. Their attenuation is proportional to the standard deviation of the autocorrelation function, just as our calculated attenuation is proportional to $\kappa^2_0$. However, they find that the attenuation is also proportional to the wavelength, which is not the case in our highly multimode ray-based picture. Another point that differentiates our work is our emphasis on refractive index-based optimization strategies to reduce the attenuation.

\noindent\textbf{Funding.} Grant number W911NF-19-1-0352 from the United States Army Research Office.

\end{document}